\documentclass{emulateapj}
\usepackage{amsmath}
 
\newcommand{\bi}[1]{\ensuremath{\boldsymbol{#1}}} 
\slugcomment{\scriptsize{Submitted to ApJ, Preprint typeset using \LaTeX style}} 
\shorttitle{Axisymmetric MRI in Viscous Accretion Disks}
\shortauthors{MASADA AND SANO}
\begin{document}
\title{Axisymmetric Magnetorotational Instability in Viscous Accretion Disks} 
\author{Youhei Masada\altaffilmark{1,2}, and Takayoshi Sano\altaffilmark{3}}
\altaffiltext{1}{Institute of Astronomy and Astrophysics, and Theoretical Institute for Advanced Research in Astrophysics, 
Academia Sinica, Taipei 10617, Taiwan, R.O.C; masada@asiaa.sinica.edu.tw}
\altaffiltext{2}{Kwasan and Hida Observatories, and Department of Astronomy, Kyoto University, Kyoto 606-8502, Japan; masada@kusastro.kyoto-u.ac.jp}
\altaffiltext{3}{Institute of Laser Engineering, Osaka University, Osaka 560-8502, Japan} 
\begin{abstract}
Axisymmetric magnetorotational instability (MRI) in viscous accretion disks is investigated by linear analysis and two-dimensional nonlinear simulations. 
The linear growth of the viscous MRI is characterized by the Reynolds number defined as $R_{\rm MRI} \equiv v_A^2/\nu\Omega $, where $v_A$ 
is the Alfv{\'e}n velocity, $\nu$ is the kinematic viscosity, and $\Omega$ is the angular velocity of the disk. Although the linear growth rate is suppressed 
considerably as the Reynolds number decreases, the nonlinear behavior is found to be almost independent of $R_{\rm MRI}$. 
At the nonlinear evolutionary stage, a two-channel flow continues growing and the Maxwell stress increases until the end of calculations 
even though the Reynolds number is much smaller than unity. A large portion of the injected energy to the system is converted to the magnetic energy. 
The gain rate of the thermal energy, on the other hand, is found to be much larger than the viscous heating rate. Nonlinear behavior of the MRI 
in the viscous regime and its difference from that in the highly resistive regime can be explained schematically by using the characteristics of 
the linear dispersion relation. Applying our results to the case with both the viscosity and resistivity, it is anticipated that the critical value 
of the Lundquist number $S_{\rm MRI} \equiv v_A^2/\eta\Omega$ for active turbulence depends on the magnetic Prandtl number 
$S_{{\rm MRI},c} \propto Pm^{1/2}$ in the regime of $Pm \gg 1$ and remains constant when $Pm \ll 1$, 
where $Pm \equiv S_{\rm MRI}/R_{\rm MRI} = \nu/\eta$ and $\eta$ is the magnetic diffusivity.
\end{abstract}
\section{Introduction}
Magnetohydrodynamic (MHD) turbulence is the most promising candidate for angular momentum transport in astrophysical disk systems. 
Nonlinear behaviors of the magnetorotational instability (MRI) are actively investigated as a driving mechanism of MHD turbulence 
over the last decades (Balbus \& Hawley 1991, 1998). The central issue in MRI research is its nonlinear properties, in particular, 
the saturation amplitude of the instability. Although the key processes governing the nonlinear saturation are scoped by global and local numerical studies, 
it is not fully explained yet (Hawley \& Balbus 1992; Hawley et al. 1995, 1996; Brandenburg et al. 1995; Matsumoto \& Tajima 1995; 
Stone et al. 1996; Hawley 2000; Machida et al. 2000; Arlt \& R\"udiger 2001; Balbus 2003). 
Recently, Lesur \& Ogilvie (2008) argue that, in the shearing box context with zero-net vertical flux, 
MHD turbulence is sustained through nonlinear classical dynamo activity once the MRI is operated. 
It would be necessary and significative to study the saturation process from the microscopic viewpoint of 
the physical sustaining mechanism for MHD turbulence as they have done. 

Ohmic dissipation is one of the crucial processes that determine the saturation amplitude of the MRI. Linear growth rate of the MRI can be reduced 
significantly because of the suppression by ohmic dissipation. Two- and three-dimensional local shearing box simulations (Sano et al. 1998, 2004; Sano \& Stone 2002) 
have shown that physical properties of the saturated turbulence depend on the Lundquist number $S_{\rm MRI } \equiv v_A^2/\eta\Omega $, 
where $v_A$ is the Alfv\'en velocity, $\Omega $ is the angular velocity, and $\eta $ is the magnetic diffusivity (see also Fleming et al. 2000; 
Ziegler \& R\"udiger 2001; Liu et al. 2006). Particularly, when $S_{\rm MRI} \ll 1$, the size of the saturated stress decreases with the decrease of $S_{\rm MRI} $ 
(Sano \& Stone 2002; Pessah et al. 2007). It is also pointed out that magnetic reconnection plays an important role in the energy dissipation of 
MRI driven turbulence (Sano \& Inutsuka 2001). 

Numerical issues are one of the main reasons why the saturation physics of the MRI is remained to be understood. 
Fromang \& Papaloizou (2007) demonstrate the efficiency of angular momentum transport decreases linearly with the grid spacing as the resolution increases. 
Although it is very difficult to distinguish between the numerical and physical factors working as the saturation mechanism (King et al. 2007; Silvers 2007), 
current researches of the MRI pay much attention to the numerical factors with the greatest care. Pessah et al. (2007) derive a scaling law of the saturated 
stress from past wide variety of numerical results by analytically eliminating the numerical factors such as box size and resolutions. 

More straightforward way for decontaminating the numerical factors is to bring explicitly the physical diffusivities much larger than the numerical one 
into the computational study. Lesur \& Longaretti (2007) have performed first systematic study of the MRI in the presence of both viscous and magnetic 
dissipations, which are larger than the numerical diffusivities. For the cases with nonzero net flux of the vertical field, the transport property 
in the saturated state depends on the magnetic Prandtl number $Pm \equiv \nu/\eta $, where $\nu $ is the kinematic viscosity. 
Fromang et al. (2007) have reported that in zero magnetic flux cases the turbulent activity is an increasing function of the magnetic Prandtl number $Pm$. 
Linear behaviors of the MRI in the presence of both the viscosity and resistivity are analytically studied by Pessah \& Chan (2008) comprehensively. 

The magnetic Prandtl number $Pm$ takes a wide range of values in astrophysical disk systems. 
In protoplanetary disks surrounding young stellar objects, the magnetic Prandtl number is much smaller than unity 
because of their low ionization degree (Nakano 1984; Umebayashi \& Nakano 1988; Sano et al. 2000). 
In accretion disks of compact X-ray sources and active galactic nuclei, the magnetic Prandtl number ranges from $\simeq 10^{-3}$ to $10^{3}$ depending 
on the distance from the central object (Balbus \& Henri 2008). In collapsar disks which is known as the central engine of gamma-ray bursts (Woosley 1993), 
the physical state with high magnetic Prandtl number of $Pm \gtrsim 10^{10}$ is expected to be realized in their evolutionary stage as a result of the 
large neutrino viscosity (Masada et al. 2007). Therefore, more systematic and deeper study on the MRI in the presence of both the viscosity and resistivity 
is quite important for understanding the accretion process triggered by the MRI in various disk systems. 

One important unsettled matter, in these situations, is the role of the kinematic viscosity at the nonlinear stage of the MRI. 
In general, the viscosity  as well as the magnetic resistivity can suppress the growth of the MRI. However the dependence of nonlinear outcome 
on the Prandtl number indicates that the role of the viscosity in MRI turbulence could be different from that of the resistivity. 
Then, the main purpose of this paper is to reveal nonlinear features of the MRI in viscous accretion disks. As is described in what follows, 
linear growth of the MRI is characterized by the Reynolds number $R_{\rm MRI} \equiv v_A^2/\nu\Omega $ in the viscous fluid, and 
by the Lundquist number $S_{\rm MRI} \equiv v_A^2/\eta\Omega $ in the resistive fluid. Focusing on these two non-dimensional parameters, 
we clarify the difference in nonlinear behaviors of the MRI between the viscous and resistive systems. 

Our paper is organized as follows. In \S~2, linear features of the MRI in the viscous fluid are presented.  In \S~3, nonlinear behavior of the MRI 
is investigated by two-dimensional MHD simulations taking account of the viscous terms. The differences between the effect of the viscosity and resistivity 
are also clarified in \S~3. Finally we make an physical explanation for our nonlinear results with the help of the linear dispersion relation. Applying our results to 
double diffusive systems, we predict a condition for sustaining active MRI turbulence in the presence of both the viscosity and resistivity in \S~4. 
\section{Linear Analysis}
First, we  provide the linear features of the MRI in a viscous accretion disk threaded by a uniform vertical field $B_z$. 
Plane-wave perturbation theory, with WKB spatial and temporal dependence $\propto \exp (i k_z z + \gamma t )$, gives a local
axisymmetric dispersion equation for the MRI,
\begin{eqnarray}
\tilde{\gamma}^4 + \frac{2}{R_{\rm MRI}}\tilde{k}_z^2\tilde{\gamma}^3 + \left[ \frac{1}{R_{\rm MRI}^2}\tilde{k}_z^4 + 2\tilde{k}_z^2 
+ \tilde{\kappa}^2 \right]\tilde{\gamma }^2 && \nonumber \\  + \frac{2}{R_{\rm MRI}}\tilde{k}_z^4\tilde{\gamma} 
+ (\tilde{k}_z^4 - 2q\tilde{k}_z^2) = 0 && \;, \label{eq1} 
\end{eqnarray}
(Menou et al. 2006; Masada et al. 2007; Pessah \& Chan 2008), where $\tilde{\gamma} = \gamma/\Omega$ is the growth rate normalized by angular velocity $\Omega$,  and $\tilde{k}_z = k_z v_{A}/\Omega$ is the vertical wavenumber normalized by $\Omega/v_{A}$. The epicyclic frequency, normalized by the angular velocity, can be  expressed as $\tilde{\kappa} = [2(2 - q)]^{1/2} $ by using the shear parameter $q \equiv - {\rm d} \ln \Omega / {\rm d} \ln r$. The Reynolds number for the MRI 
is defined as $R_{\rm MRI} = VL/\nu \equiv v_A^2/\nu\Omega$. Here the
characteristic velocity and length are $V = v_A$ and $L = v_A/\Omega
$,  respectively. In this paper, we focus on the MRI in Keplerian
disks where the epicyclic frequency is equal to the angular velocity
($q = 3/2$). 

\begin{figure}
\begin{center}
\scalebox{2.0}{\rotatebox{0}{\includegraphics{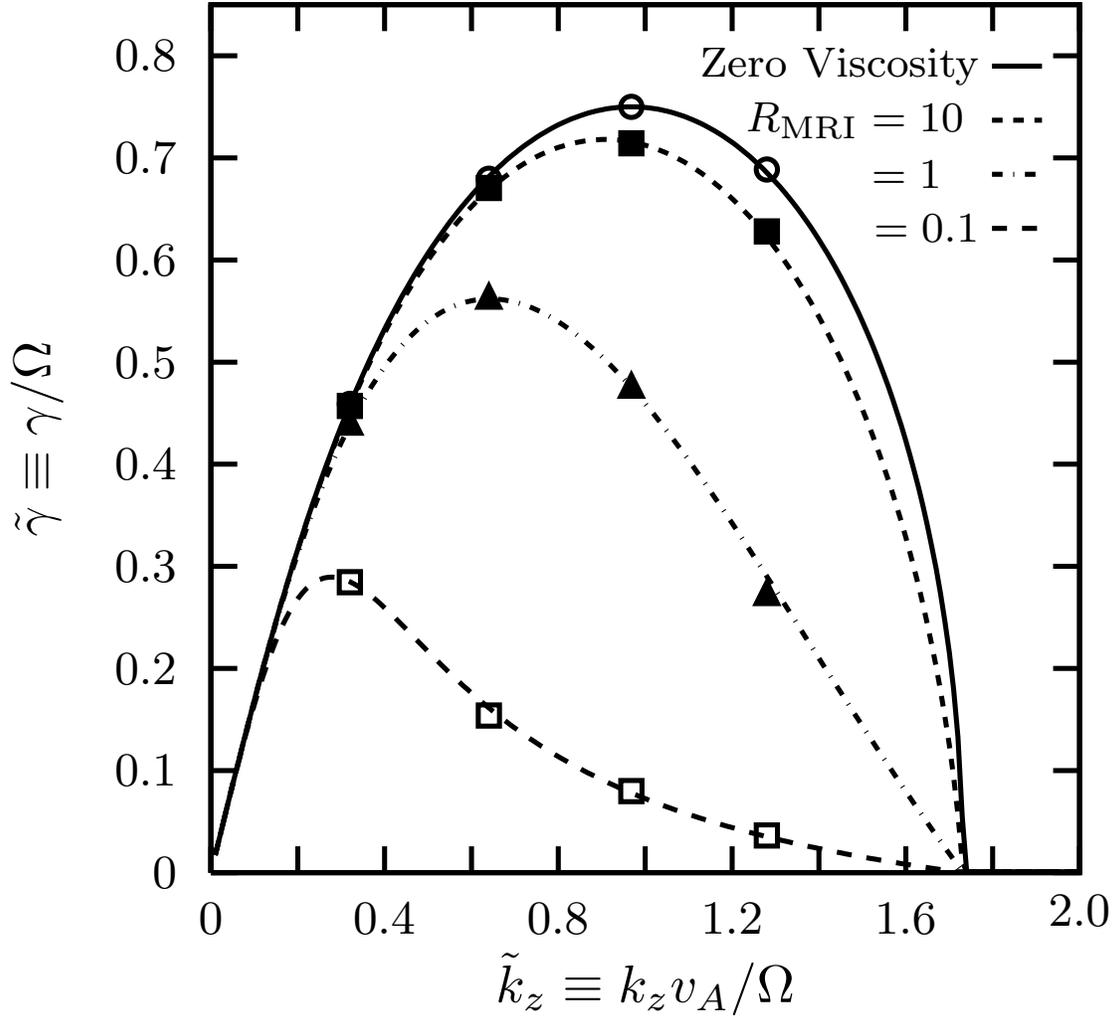}}} \\
\caption{Linear growth rate of the MRI as a function of the wavenumber. The cases with different Reynolds numbers for MRI 
$R_{\rm MRI} \equiv v_A^2/\nu\Omega = \infty$, $10$, $1.0$ and $0.1$ are depicted. Normalization of the vertical and horizontal axes are 
the angular velocity $\Omega$ and the typical wavenumber of the MRI $v_A/\Omega$, respectively. 
The symbols plotted over the analytical dispersion relation are numerical growth rates calculated by our numerical scheme. }
\label{fig1}
\end{center}
\end{figure}

The dispersion equation~(\ref{eq1}) is characterized by the Reynolds number $R_{\rm MRI}$. 
The linear growth rate of the MRI is shown as a function of the vertical wavenumber for the cases $R_{\rm MRI} = 0.1$, $1.0$, $10.0$ and $\infty $ in Figure~\ref{fig1}.  When the kinematic viscosity is negligible ($R_{\rm MRI} \gg 1$), the dispersion relation is identical to that of the ideal MHD case. 
If the Reynolds number is less than unity, the growth of the MRI is suppressed and the maximum growth rate is reduced significantly. 
The most unstable wavenumber decreases with decreasing $R_{\rm MRI}$, because the viscous damping becomes more efficient 
for shorter wavelength perturbations. However, it is interesting that the critical wavenumber for the instability, $\tilde{k}_{z, {\rm crit}} = \sqrt{2q}$, 
remains unchanged in spite of the size of $R_{\rm MRI}$. 

\begin{figure}
\begin{center}
\rotatebox{-90}{
\begin{tabular}{cc}
\scalebox{1.3}{{\includegraphics{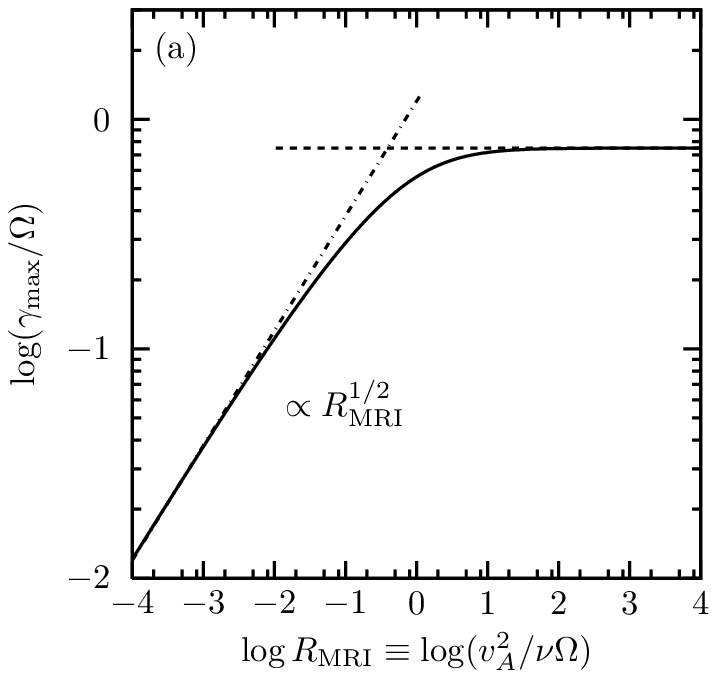}}} &
\scalebox{1.3}{{\includegraphics{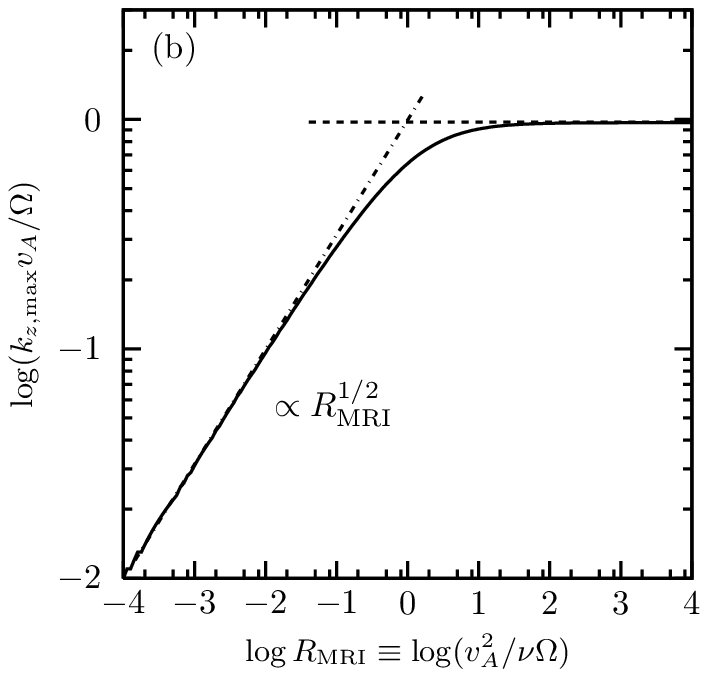}}} 
\end{tabular}
}
\caption{(a) The maximum growth rate normalized by the angular velocity $\Omega $ and (b) the fastest growing wavenumber normalized by $v_A/\Omega $ 
are shown as functions of the Reynolds number $R_{\rm MRI}$. In the
range $R_{\rm MRI} \ll 1$, both quantities are proportional to the
square root of the Reynolds number $R_{\rm MRI}^{1/2}$.  } 
\label{fig2}
\end{center}
\end{figure}

Figures~\ref{fig2}a and \ref{fig2}b show the growth rate and wavenumber of the fastest growing mode as a function of the Reynolds number. 
These figures indicate that $\tilde{\gamma}_{\rm max}$ and $\tilde{k}_{z,{\rm max}}$ are proportional to $R_{\rm MRI}^{1/2}$ 
when $R_{\rm MRI} \ll 1$ (Pessah \& Chan 2008). 
In the regime of $R_{\rm MRI} \ll 1$, the dispersion equation~(\ref{eq1}) can be simplified and reduced to 
\begin{equation}
[K^4 + 2(2 - q) ]\ G^2 - 2q K^2 = 0 \;, \label{eq2}
\end{equation}
where $G \equiv \tilde{\gamma}/R_{\rm MRI}^{1/2}$ and $K \equiv \tilde{k}/R_{\rm MRI}^{1/2}$. 
The fastest growing wavelength and the maximum growth rate are obtained analytically from this equation; 
\begin{eqnarray}
\tilde{k}_{z,{\rm max}} & \equiv & \left( \frac{k_{z,{\rm max}} v_A}{\Omega }\right) = [2(2 - q)]^{1/4}\ R_{\rm MRI}^{1/2} \;, \label{eq3} \\ 
\tilde{\gamma}_{\rm max} & \equiv & \left( \frac{\gamma_{\rm max}}{\Omega} \right) = \left[ \frac{q^2}{2(2-q)} \right]^{1/4} \ R_{\rm MRI}^{1/2}\;. \label {eq4} 
\end{eqnarray} 

Qualitative features of the fastest growing mode can be explained as follows: Keplerian shear flow is a key ingredient of the unstable growth of the MRI.
The viscous dissipation affects the growth of the MRI when the damping rate is comparable to the shear rate for a perturbation. 
Then the viscous damping rate of the unstable mode would be balanced with the shear rate in the viscous regime, 
 $k_{z}^2\nu \simeq | {\rm d}\Omega/{\rm d}\ln r| \simeq \Omega $. This relation gives the most unstable wavenumber $k_{z,{\rm max}} \simeq (\Omega / \nu)^{1/2}$, 
and thus $\tilde{k}_{z,{\rm max}} \propto R_{\rm MRI}^{1/2}$. Since the maximum growth rate of the MRI is equal to the Alfv\'en frequency of the fastest growing mode, 
it corresponds to $\gamma_{\max} \simeq k_{z,{\rm max}} v_A$, or $\tilde{\gamma}_{\rm max} \propto R_{\rm MRI}^{1/2}$. This is why the normalized wavenumber 
and growth rate of the fastest growing mode are proportional to the $R_{\rm MRI}^{1/2}$ in the presence of large viscous dissipation. 

Note that the $R_{\rm MRI}$-dependence of the fastest growing mode is slightly different from that on the Lundquist number 
($S_{\rm MRI} \equiv v_A^2/\eta\Omega $) in the presence of ohmic dissipation (Sano \& Miyama 1999), where $\eta $ is the magnetic diffusivity. 
Based on the linear analysis, the resistivity can suppress the MRI more efficiently compared to the viscosity. In the resistive regime, the ohmic dissipation 
can suppress the MRI when the dissipation time of the unstable mode is comparable to the Alfv\'en time, $\lambda_{\rm max}^2/\eta \simeq \lambda_{\rm max}/v_A$, 
Thus the most unstable wavelength is given by $\lambda_{\max} \simeq \eta / v_A$. The growth rate of the fastest growing mode is given by 
$\gamma_{\rm max} \simeq v_A / \lambda_{\max} \simeq v_A^2 / \eta$. Then we can obtain the relations $\tilde{k}_{z,{\rm max}} \propto S_{\rm MRI}$ 
and $\tilde{\gamma}_{\rm max} \propto S_{\rm MRI}$. It is stressed that the critical wavenumber $k_{z,{\rm crit}}$ in the resistive regime is also 
proportional to the Lundquist number, $\tilde{k}_{z,{\rm crit}} \propto S_{\rm MRI}$. 
Unstable wavelength and growth rate of the MRI for ideal MHD, resistive, and viscous cases are summarized in Table~\ref{table1}.  
\section{Nonlinear Analysis}
\subsection{Numerical Setting}
For elucidating the nonlinear features of the MRI, viscous MHD equations are solved with a finite-differencing code 
which was developed by Sano et al. (1998). The hydrodynamic module of our scheme is based on the second-order Godunov scheme 
(van Leer 1979), which consists of Lagrangian and remap steps. The Riemann solver is modified for accounting the effect of tangential
magnetic fields. The field evolution is calculated with the Consistent MoC-CT method (Clarke 1996). 
The energy equation is solved in the conservative form and the viscous terms are calculated in the Lagrangian step. 
The advantages of our scheme are its robustness for strong shocks and the satisfaction of the divergence-free 
constraint of magnetic fields (Evans \& Hawley 1988; Stone \& Norman 1992). 

We use a local shearing box model (Hawley et al. 1995). In this approximation, equations of viscous MHD are 
written in a local Cartesian frame of reference corotating with the disk at the angular velocity $\Omega $ corresponding to a fiducial radius $R$. 
Then the coordinates are presented as $x = r - R $, $y = R\phi - \Omega t$, and $z$. The fundamental equations are written in terms of 
these coordinates within a small region surrounding the fiducial radius, in $\Delta r \ll R $,  
\begin{eqnarray}
& & \frac{\partial \rho}{\partial t} + \nabla\cdot (\rho \bi{v})   =  0 \;, \label{eq7} \\ 
\frac{{\partial }\bi{v}}{ \partial  t} + \bi{v}\cdot \nabla \bi{v}  & = & - \frac{1}{\rho} \nabla P_{\rm eff}+ \frac{(\bi{B}\cdot \nabla)\bi{B}}{4\pi\rho }  \nonumber \\ 
&& - 2\Omega\times \bi{v} + 2q\Omega^2 x {\tilde{\bi{x}}} + \bi{R}  \;, \label{eq8} \\
\frac{\partial \epsilon }{\partial t} + (\bi{v}\cdot\nabla) \epsilon & = & - \frac{P}{\rho} \nabla\cdot \bi{v} + \Phi \;, \label{eq9}
\end{eqnarray}
\begin{equation}
\frac{\partial \bi{B} }{\partial t} = \nabla \times (\bi{v} \times \bi{B} ) \;, \label{eq10}
\end{equation}
where
\begin{eqnarray}
P_{\rm eff}  & = & P + \frac{| \bi{B}|^2}{8\pi} - \rho \xi(\nabla\cdot\bi{v}) \;, \label{eq11} \\
R_{ij} & = & \frac{1}{\rho}\frac{\partial}{\partial x_{i}}\left[\rho\nu \left(\frac{\partial v_{i} }{\partial x_{j} } + 
\frac{\partial v_{j}}{\partial x_{i}} \right)\right] \;, \label{eq12} \\
\Phi_{ij} & = &  \xi (\nabla\cdot {\boldmath v})^2 
+ \frac{\nu}{2} \left( \frac{\partial v_i}{\partial x_j} + \frac{\partial v_j}{\partial x_i} \right)^2 \;. \label{eq13} 
\end{eqnarray}
where $\epsilon $ is the specific internal energy, $\xi \equiv \chi - 2\nu /3$, and  $\chi$ is the bulk viscosity. 
In the following, the effect of the bulk viscosity is neglected ($\chi \ll \nu $). The term $2q\Omega^2 x$ in the momentum equation 
is the tidal expansion of the effective potential. Assuming the ideal gas, the pressure is given by $P = (\gamma -1)\rho \epsilon $. 
The spatially uniform kinematic viscosity and a constant ratio of specific heats ($\gamma = 5/3$) are considered. 

Since we focus on the local properties of the instability, we employ a numerical grid representing a small section of the disk interior for a local disk model. 
Adopting the Keplerian shear flow as unperturbed state, the azimuthal velocity is given by $v_y = -q\Omega x$ in the frame corotating with the velocity $R\Omega$. 
The initial field geometry is a weak uniform field in the vertical direction $B_z = B_0$. The radial force balance at the initial state is thus realized between 
the Coriolis force and the tidal force. 

Two-dimensional calculation is performed in the radial-vertical plane with a volume bounded by $x = z = \pm H/2$, 
where $H\equiv (2/\gamma )^{1/2} c_s/\Omega $ is the scale height of the disk. We use a uniform grid of $128\times 128$ zones. 
A periodic boundary condition is applied in the vertical direction. For the radial boundary condition, we adopt a sheared periodic boundary condition 
(Hawley et al. 1995). In this model, the vertical component of gravity can be ignored. Except for the shear velocity, 
the physical quantities are thus assumed to be spatially uniform; $\rho = \rho_0$ and $P = P_0$ where $\rho_0$ and $P_0$ are constant values. 
We choose normalizations with $\rho_0 = 1$, $H=1$, $\Omega = 10^{-3}$, and $P_0 = 5\times 10^{-7} $.
Initial perturbations are introduced as spatially uncorrelated velocity and adiabatic pressure fluctuations. 
These fluctuations have a zero mean value with a maximum amplitude of $| \delta P|/P_0 = 10^{-2} $ and $|\delta \bi{v}|/c_s = 10^{-2}$. 

Our local disk model is characterized by non-dimensional parameters, which are the plasma beta of the initial field strength $\beta_0 = 8\pi P_0/B_0^2$ 
and the initial Reynolds number $R_{\rm MRI}$. In this paper, we show the results focusing on the effects of the Reynolds number. 
In what follows, we fix the initial field strength as $\beta_0 = 10^4$ in all our models. In the case $R_{\rm MRI} \ll 1$, 
the fastest growing wavelength of the local disk system $\tilde{\lambda}_{\rm vis}$ can be defined from equation~(\ref{eq3}), 
\begin{equation}
\tilde{\lambda}_{\rm vis}  \equiv  \frac{\lambda_{\rm max} }{H} = \frac{2\pi}{H} \left( \frac{\nu }{\Omega} \right)^{1/2} = \frac{2\pi }{\sqrt{\beta R_{\rm MRI}}}  \;, \label{eq14}
\end{equation}
where $\lambda_{{\rm max} }$ is the fastest growing wavelength expected by the linear analysis for viscous MHD case. 
To capture the most unstable mode in the computational domain, $\tilde{\lambda}_{\rm vis} \lesssim 1$, the Reynolds number must be $R_{\rm MRI} \gtrsim 0.004$. 
\subsection{Results}
\subsubsection{Growth Rate at the Linear Phase}
We investigate the growth rate at the early linear phase in order to confirm the accuracy of our numerical scheme. All the simulations 
begin with random perturbation of very small amplitude so that any
growing modes should be well described by a linear analysis during
the first few orbits 
of the evolution. The time history of each mode is followed through a two-dimensional Fourier decomposition carried out 
at frequent time intervals. We define the Fourier coefficient $a_k$ of the radial velocity $v_x$ as 
\begin{equation}
a_k (t) = \frac{1}{(2\pi )^2}\int\int v_x (x,z,t) e^{ik_z z} dxdz \;. \label{eq15}
\end{equation} 
Here we select the modes with $k_x = 0$ because they are the most unstable. 

The numerical growth rates in early linear phase are plotted over the analytic dispersion relation in Figure~\ref{fig1}. The growth rate at 
$\tilde{k}_z = 0.32$, $0.64$, $0.97$, and $1.28$ are shown as representative cases, which are obviously reproducing the analytic results for all the cases. 
Thus our numerical scheme has ability to simulate correctly the unstable growth of perturbations excited by the MRI in the presence of the viscous dissipation. 
\subsubsection{Dependence on the Reynolds Number}
The efficiency of angular momentum transport is given by the $x$-$y$ component of the stress tensor, 
\begin{equation}
w_{xy}   = w_{M} + w_{R} = - \frac{ B_x B_y }{ 4\pi} +  \rho v_x \delta v_y \;,  \label{eq16}
\end{equation}
where $w_{M} $ and $w_{R}$ are Maxwell and Reynolds stresses, respectively. This is related to the $\alpha$ parameter of Shakura \& Sunyaev (1973) 
by $\alpha \equiv w_{xy}/P = (w_{M} + w_{R})/P$. The Maxwell stress is proportional to the magnetic energy and usually dominates over the Reynolds stress 
in MRI driven turbulence. 

\begin{figure}
\begin{center}
\rotatebox{-90}{
\begin{tabular}{cc}
\scalebox{1.3}{{\includegraphics{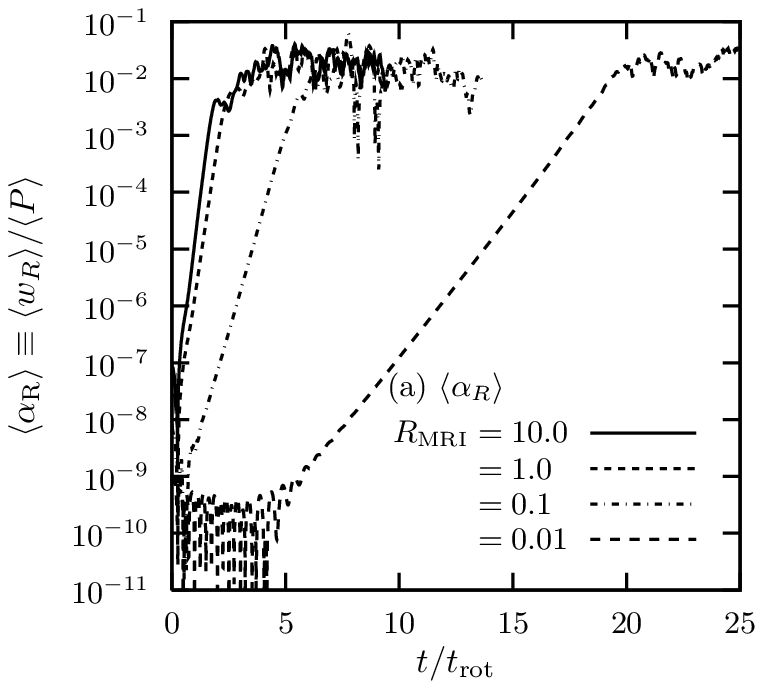}}} &
\scalebox{1.3}{{\includegraphics{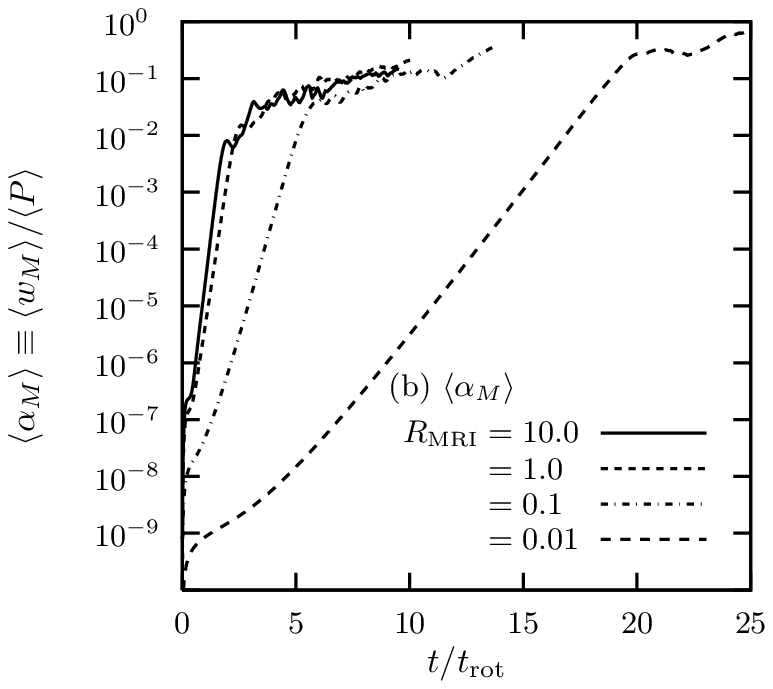}}} 
\end{tabular}}
\caption{Time evolution of the volume-averaged (a) Reynolds stress $\langle \alpha_{R} \rangle \equiv \langle w_R \rangle / \langle P \rangle$ and 
(b) Maxwell stress $\langle \alpha_{M} \rangle \equiv \langle w_M \rangle / \langle P \rangle$ for the cases with different Reynolds number 
$R_{\rm MRI} = 10.0$, $1.0$, $0.1$ and $0.01$. The horizontal axis is normalized by the disk rotation time $t_{\rm rot} = 2\pi/\Omega $. } 
\label{fig3}
\end{center}
\end{figure}

To demonstrate the nonlinear features of the MRI, the time-evolutions of volume-averaged Reynolds and Maxwell stresses, 
$\langle \alpha_{R} \rangle \equiv \langle w_R \rangle / \langle P \rangle$ and $\langle \alpha_{M} \rangle \equiv \langle w_M \rangle / \langle P \rangle$,
are depicted in Figure~\ref{fig3} for the cases with different Reynolds numbers $R_{\rm MRI} = 0.01$, $0.1$, $1.0$, and $10.0$. The single bracket indicates 
a volume average of physical quantities. The horizontal axis is the time normalized by the rotation time $t_{\rm rot} \equiv 2\pi/\Omega $. 
The kinematic viscosity in those models is equivalent to the $\alpha$ parameter of the size $\alpha_{\nu} \simeq (R_{\rm MRI} \beta_0)^{-1} \lesssim 0.01$. 

The linear growth rate decreases as the Reynolds number decreases. After the linear growth of the MRI, a two-channel flow appears for the ideal MHD cases 
(Hawley \& Balbus 1992). The two-channel flow is an axisymmetric MRI mode whose vertical wavelength fits the vertical box size. 
This linearly unstable mode is also an exact solution of nonlinear MHD equations, so that the magnetic field can be amplified exponentially 
even at the nonlinear regime (Goodman \& Xu 1994). Similar behavior is found in all the viscous models even though the Reynolds number is much 
smaller than unity. The magnetic energy continues growing and is not saturated even at the nonlinear regime. The Maxwell stress increases until 
the end of calculations for all the models. The Reynolds stress, on the other hand, approaches a constant value at the nonlinear stage. 

\begin{figure}
\begin{center}
\rotatebox{-90}{
\begin{tabular}{cc}
\scalebox{1.3}{{\includegraphics{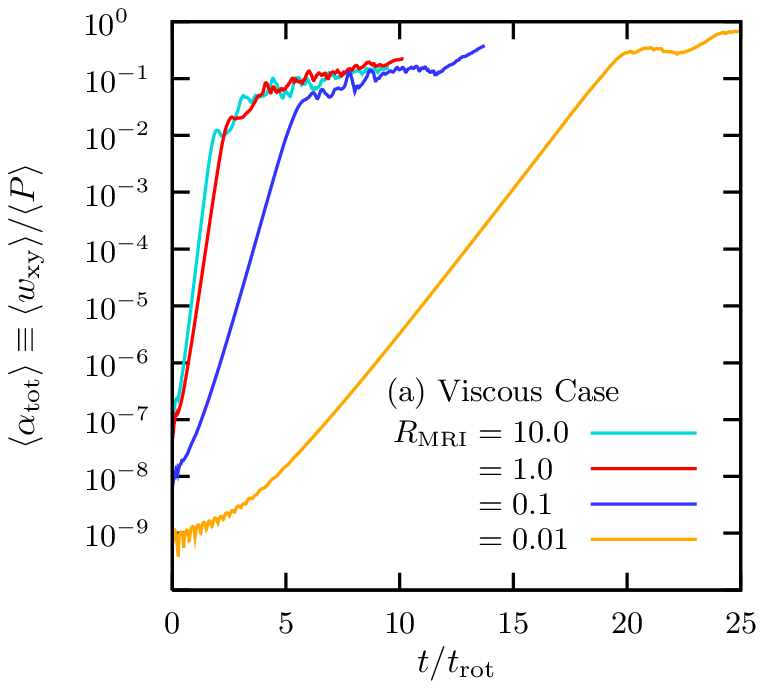}}} &
\scalebox{1.3}{{\includegraphics{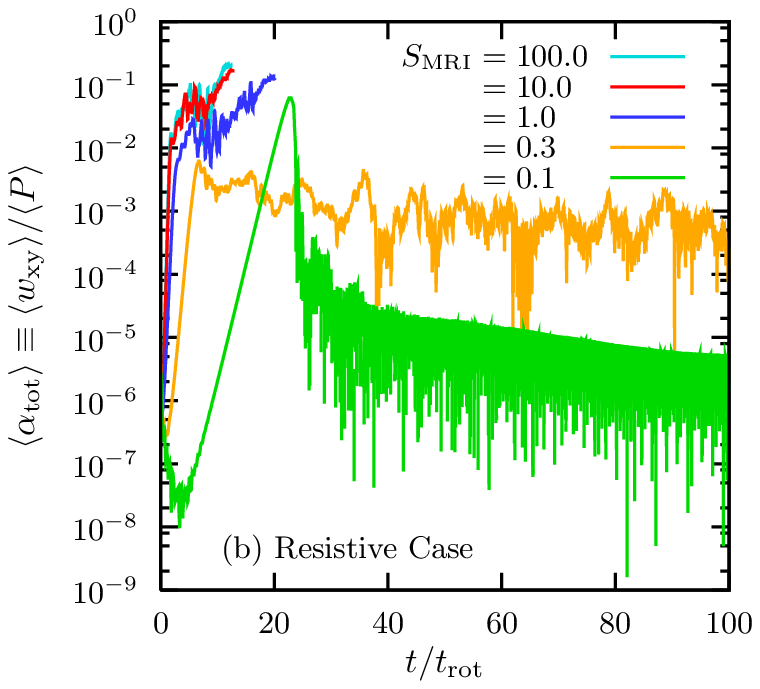}}} 
\end{tabular}
}
\caption{Panel~(a): Time evolution of the volume-averaged $\alpha$ parameter of Shakura \& Sunyaev (1973) in the viscous fluid for the cases with 
different Reynolds number $R_{\rm MRI} = 10.0$, $1.0$, $0.1$ and
$0.01$. Panel~(b): The $\alpha$ parameter in the resistive fluid for the cases with different
Lundquist number $S_{\rm MRI} = 100.0$, $10.0$, $1.0$, $0.3$ and $0.1$. Normalization of the horizontal axis is the same as Figure~\ref{fig3}. } 
\label{fig4}
\end{center}
\end{figure}

The time evolution of $\langle \alpha_{\rm tot} \rangle$, which is the sum of $\langle \alpha_{R} \rangle $ and $\langle \alpha_{M} \rangle $, is shown 
in Figure~\ref{fig4}a. The nonlinear behavior of the MRI in viscous fluid is quite different from that in the models taking account of the magnetic diffusivity 
(Sano et al. 1998, 2004; Fleming et al. 2000; Sano \& Inutsuka 2001). For the purpose of comparison, Figure~\ref{fig4}b shows $\langle \alpha_{\rm tot} \rangle$ 
in the resistive MHD runs with different Lundquist numbers $S_{\rm MRI}$. The initial conditions are the same as the viscous models except for the
dissipation terms. When the dissipation processes can be negligible ($R_{\rm MRI} \gtrsim 1$ and $S_{\rm MRI} \gtrsim 1$), the evolution is quite similar to 
that of ideal MHD. The viscous dissipation cannot damp MRI driven turbulence in the nonlinear regime even though $R_{\rm MRI} \ll 1$. 
In contrast, the MRI saturates and MHD turbulence dies away at the nonlinear regime for the case with $S_{\rm MRI} \lesssim 1$. 
\subsubsection{Energy Injection into the Local System}
The total energy within the shearing box is defined as $\Gamma \equiv \int {\rm d} V [ \rho(v^2/2 + \epsilon +\phi) + B^2/8\pi ] $, where 
$\phi = -q\Omega^2x^2$ is the tidal expansion of the effective potential (Hawley et al. 1995). Using the evolution equations for viscous MHD system 
[eqs.~(\ref{eq7})--(\ref{eq13})], the time derivative of the total energy gives 
\begin{eqnarray}
\frac{{\rm d} \Gamma }{{\rm d} t} & = & q\Omega L_x \int_x {\rm d} A 
\left[ \left( \rho v_x \delta v_y - \frac{B_xB_y}{4\pi} \right) - \rho\nu \frac{\partial v_y}{\partial x} \right] \nonumber \\
& \simeq & q\Omega L_x \left( \int_x {\rm d}A\ w_{xy} + \nu q \Omega \int_x {\rm d}A\  \rho \right) \;, \label{eq17}
\end{eqnarray}
where d$A$ is the surface element and the integral is taken over either of the radial boundaries. We can derive the last term of above 
equation by assuming that the radial gradient of the perturbed azimuthal velocity is negligible, that is $\partial v_y /\partial x \simeq - q \Omega$. 
The energy injection rate through the radial boundary is proportional to the kinematic viscous stress as well as the turbulent stress $w_{xy}$ at the boundary. 

Using the volume-averaged values instead of the surface-averaged ones at the radial boundary, 
the volume average of the total energy changing rate $\langle \dot{E}_{\rm tot} \rangle $ is given by 
\begin{equation} 
\langle \dot{E}_{\rm tot} \rangle =  \langle \dot{E}_{{\rm in},w} \rangle + \langle \dot{E}_{{\rm in},v} \rangle \;, \label{eq18}
\end{equation} 
where $\langle \dot{E}_{{\rm in},w} \rangle \equiv q\Omega \langle w_{xy} \rangle $ is the energy injection rate caused by the turbulent stress 
(Sano \& Inutsuka 2001), and $\langle \dot{E}_{{\rm in},v} \rangle \equiv \nu q^2\Omega^2 \langle \rho \rangle $ is done by the kinematic viscous stress. 
The second term is always positive. When the turbulent stress is positive, the total energy of the system must increase. The source of the input energy 
is the background shear motion. In realistic disk systems, positive stresses lead to inward mass accretion, bringing a loss of gravitational energy. 
The gain in total energy in the shearing box represents this energy release. 

\begin{figure}
\begin{center}
\scalebox{2.0}{{\includegraphics{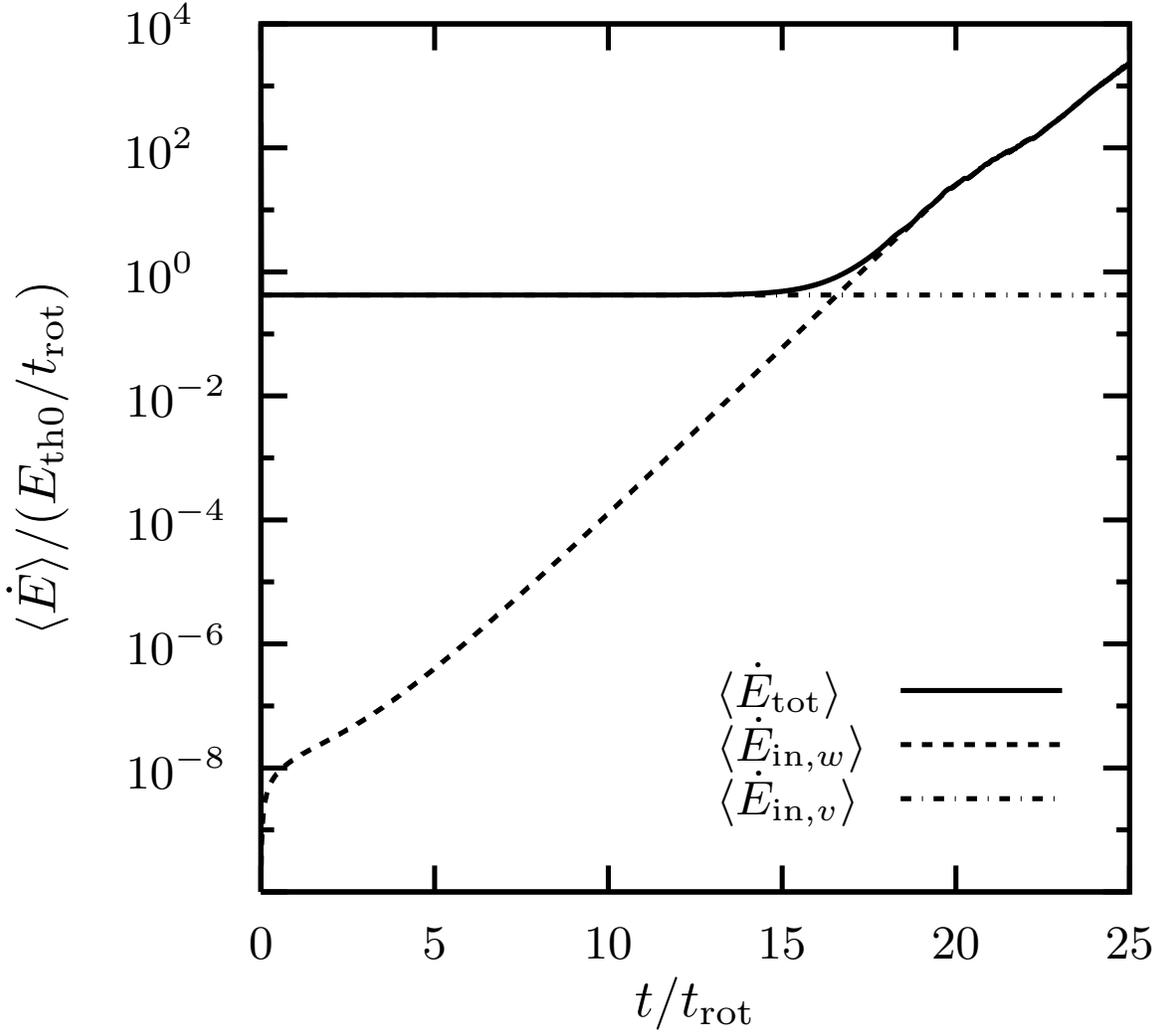}}}
\caption{Time evolution of the volume-averaged time derivative of the total energy 
$\langle \dot{E}_{\rm tot} \rangle \equiv \langle \dot{E}_{\rm th} \rangle + \langle \dot{E}_{m} \rangle + \langle \dot{E}_k \rangle $ 
and the input energies due to the turbulent stress $\langle \dot{E}_{{\rm in},w} \rangle$ and the kinematic 
viscous stress $\langle \dot{E}_{{\rm in},v} \rangle $. These quantities should satisfy the energy conservation 
$\langle \dot{E}_{\rm tot} \rangle = \langle \dot{E}_{{\rm in},w} \rangle + \langle \dot{E}_{{\rm in},v} \rangle $. 
This is for the case with $R_{\rm MRI} = 0.01$. The vertical axis is given in the unit of $E_{\rm th0}/t_{\rm rot}$ 
where $E_{\rm th0} = P_0/(\gamma -1)$ is the initial thermal energy.} 
\label{fig5}
\end{center}
\end{figure}

The energy budget in our simulations can be satisfying the relation~(\ref{eq18}), because our numerical scheme solves the energy equation 
in terms of the total energy. Figure~\ref{fig5} shows the time evolution of $\langle \dot{E}_{\rm tot} \rangle $, $\langle \dot{E}_{{\rm in},w} \rangle $ 
and $\langle \dot{E}_{{\rm in},v} \rangle $ for the case with $R_{\rm MRI} = 0.01$. The vertical axis is given in the unit of $E_{\rm th0}/t_{\rm rot}$, 
where $E_{\rm th0} = P_0/(\gamma -1)$ is the initial thermal energy. During the early linear phase until about 15 orbits, the kinematic viscous stress 
takes a major role in the energy injection. In contrast, the turbulent stress becomes predominant after the MRI grows sufficiently. 
The sum of these contributions is exactly identical to the energy gain of the system in our simulations.  
\subsubsection{Energy Dissipation at the Nonlinear Stage}
\begin{figure}
\begin{center}
\rotatebox{-90}{
\begin{tabular}{cc}
\scalebox{1.2}{{\includegraphics{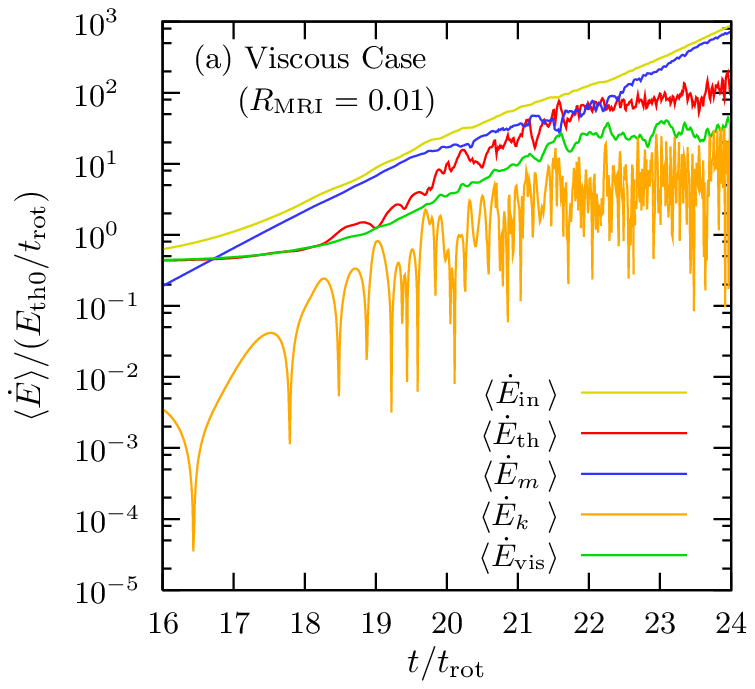}}} &
\scalebox{1.2}{{\includegraphics{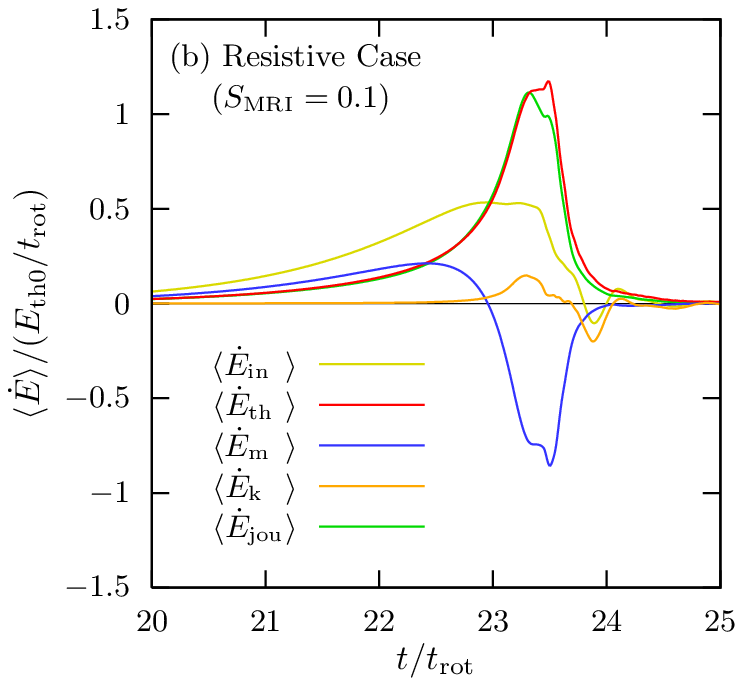}}} 
\end{tabular}
}
\caption{Time evolution of the volume-averaged time derivative of the input energy $\langle \dot{E}_{\rm in} \rangle = \langle \dot{E}_{{\rm in},w} \rangle + 
\langle \dot{E}_{{\rm in},v} \rangle$, thermal energy $\langle \dot{E}_{\rm th} \rangle$, magnetic energy $\langle \dot{E}_{m} \rangle $, kinetic energy 
$\langle \dot{E}_{k} \rangle$. Panel~(a) shows the result of the viscous case with $R_{\rm MRI} = 0.01$ in the period $16 \le t/t_{\rm rot} \le 25$. 
The volume-averaged viscous heating rate $ \langle \dot{E}_{\rm vis} \rangle$ is shown in this figure.
Panel~(b) depicts that of the resistive case with $S_{\rm MRI} = 0.1$ in the period $20 \le t/t_{\rm rot} \le 25$. 
The volume-averaged joule heating rate $ \langle \dot{E}_{\rm jou} \rangle$ is  shown in this figure.
Normalizations of each axis are the same as those in Figure~\ref{fig4}. }
\label{fig6}
\end{center}
\end{figure}

The role of dissipation processes in the energy conversion is investigated in this subsection. 
The injected energy $\langle \dot{E}_{\rm in} \rangle = \langle \dot{E}_{{\rm in},w} \rangle + \langle \dot{E}_{{\rm in}, v} \rangle $ should be balanced 
with the increase of the sum of $ {E}_{\rm th} = P/(\gamma -1)$, ${E}_{\rm m} = B^2/8\pi $, and $ {E}_{\rm k}  = \rho v^2/2$. Figure~\ref{fig6}a depicts the 
time evolution of $ \langle \dot{E}_{\rm th} \rangle $, $\langle \dot{E}_{\rm m} \rangle $, and $\langle \dot{E}_{\rm k}  \rangle $ for the case with $R_{\rm MRI} = 0.01$. 
The viscous heating rate $\langle \dot{E}_{\rm vis} \rangle$ is also shown in this figure, which is defined as
\begin{equation}
\dot{E}_{\rm vis} = \Phi_{ij} = \xi (\nabla\cdot {\boldmath v})^2  + \frac{\nu}{2} \left( \frac{\partial v_i}{\partial x_j} + \frac{\partial v_j}{\partial x_i} \right)^2 \;, \label{eq19}
\end{equation}
where $\Phi_{ij}$ is the same definition that is used in the energy equation~(\ref{eq13}). 

The sum of the components perfectly coincides with the energy gain of the system.  
Using the time- and volume-averaged values, which are indicated by double brackets, the gain rates of each energy component are 
$\langle\langle \dot{E}_{\rm th} \rangle\rangle / \langle\langle \dot{E}_{\rm in } \rangle\rangle = 0.233$, 
$\langle\langle \dot{E}_{\rm m} \rangle\rangle / \langle\langle \dot{E}_{\rm in} \rangle\rangle = 0.757$, and 
$\langle\langle \dot{E}_{\rm k}  \rangle\rangle / \langle\langle \dot{E}_{\rm in} \rangle\rangle \lesssim 0.01$. 
Here the time average is taken at $18 \le t/t_{\rm rot} \le 25$. This indicates that a large portion of the injected energy 
is converted into the magnetic energy of the system. 

Because the system is almost incompressible, the dominant heating
mechanism in our simulations should be 
the kinematic viscous heating. However the thermal energy gain is much larger than the viscous heating rate (see Fig.~\ref{fig6}a). 
The ratio of these components is evaluated as $\langle\langle \dot{E}_{\rm vis} \rangle\rangle/\langle\langle \dot{E}_{\rm th} \rangle\rangle = 0.36$ 
in the range $18 \le t/t_{\rm rot} \le 25$. This fraction becomes smaller and smaller for the cases with larger Reynolds number. 
The rest of the heating is caused by the numerical magnetic dissipation. Then our results suggest that the magnetic dissipation 
is the dominant mechanism of heating at the nonlinear phase of the MRI and might play an essential role for the suppression of MRI driven turbulence. 

Actually in the resistive MHD cases, the heating of the system is controlled by the joule heating almost completely (Sano \& Inutsuka 2001). 
The energy changing rates in a very resistive model with the Lundquist number $S_{\rm MRI} = 0.1$ is demonstrated in Figure~\ref{fig6}b. 
This is a typical case in which the channel flow is disrupted and MHD turbulence is damped at the nonlinear stage (see Fig.~\ref{fig4}a). 
The ratio of the total energy gain and the thermal energy gain is $\langle\langle \dot{E}_{\rm th} \rangle\rangle / \langle\langle \dot{E}_{\rm in } \rangle\rangle = 0.99$ 
in the range $20 \le t/t_{\rm rot} \le 25$. We find that the joule heating takes a major role in the thermal energy gain of the system, that is 
$\langle\langle \dot{E}_{\rm jou} \rangle\rangle / \langle\langle \dot{E}_{\rm th } \rangle\rangle = 0.93$.
In the resistive system, the magnetic energy amplified by the MRI is transformed into the thermal energy via joule heating or 
magnetic reconnection throughout the nonlinear evolution.   
\subsubsection{Stress at the Nonlinear Stage}
Finally, the time- and volume-averaged $\alpha_{\rm tot} $ at the nonlinear stage is depicted as a function of the 
initial Reynolds number $R_{\rm MRI}$ and Lundquist number $S_{\rm MRI}$ in Figure~\ref{fig7}. We take the average of $\langle \alpha_{\rm tot} \rangle $ over 
5 orbits just after the time when the ratio $\langle \dot{E}_{\rm th} \rangle / \langle \dot{E}_{\rm m} \rangle $ begins to rise and the nonlinear evolution is started. 
Diamond-shape shows the results in the viscous fluid and cross-shape is that in the resistive one. Note that these are not the saturated values. 
Upward arrow over-plotted on the symbols denotes that the value is the lower limit, because $\langle\langle \alpha_{\rm tot} \rangle\rangle $
is still increasing with time. The downward arrow stands for decaying models and thus the stress is the upper limit. 

The stress at the nonlinear stage are almost the same when the diffusion is weak ($R_{\rm MRI} \gtrsim 1$ or $\ S_{\rm MRI} \gtrsim 1$). 
However, a huge difference can be seen in the highly diffusive regime. In the presence of the ohmic dissipation, the stress rapidly decreases with decreasing 
$S_{\rm MRI}$. For the models with the kinematic viscosity, on the other hand, it increases with the decrease of $R_{\rm MRI}$. 
The origin of this difference between viscous and ohmic dissipative systems is discussed later in \S~4.1. 

The inverse correlation between $\langle \langle \alpha_{\rm tot} \rangle \rangle $ and $R_{\rm MRI}$ for the cases with large viscosity could be originated 
from stable growth of a channel flow. This is because the large viscosity can suppress the growth of any other modes than the two-channel flow, 
and thus the channel mode can evolve up to highly nonlinear amplitude. The viscosity may enhance the saturation amplitude of the MRI. 
However, our results are restricted in two-dimensional simulations. The nonlinear evolution of the MRI in three-dimension must be quite different, 
because the channel flow is known to be unstable to the nonaxisymmetric parasitic instability (Goodman \& Xu 1994). 
We are planning to perform three-dimensional study of the MRI for verifying these nonlinear properties in the viscous accretion disks.
\section{Discussion} 
\subsection{Nonlinear Behavior in the Single Diffusive Systems} 
To give a physical explanation for the nonlinear behavior of the axisymmetric MRI, we focus on the critical wavelength obtained from the linear theory in this section.
The diagrams of Figures~\ref{fig8}a and \ref{fig8}b indicate the critical and the fastest growing wavelengths of the MRI as a function of the Lundquist number 
$S_{\rm MRI}$ and the Reynolds number $R_{\rm MRI}$, respectively (see also Table~\ref{table1}). 
Note that the vertical axes are normalized by $2\pi (\eta/\Omega)^{1/2} $ in Figure~\ref{fig8}a and by $2\pi (\nu/\Omega)^{1/2}$ in Figure~\ref{fig8}b. 
Shaded area denotes the linearly unstable regions for the MRI.  Assuming fixed diffusivities, $S_{\rm MRI}$ and $R_{\rm MRI}$ increase as the instability 
grows because they are proportional to the squared Alfv\'en velocity. 
Then the horizontal axis in Figure~\ref{fig8} can be regarded as the time direction in terms of the evolution of the MRI.

\begin{figure}
\begin{center}
\scalebox{1.8}{{\includegraphics{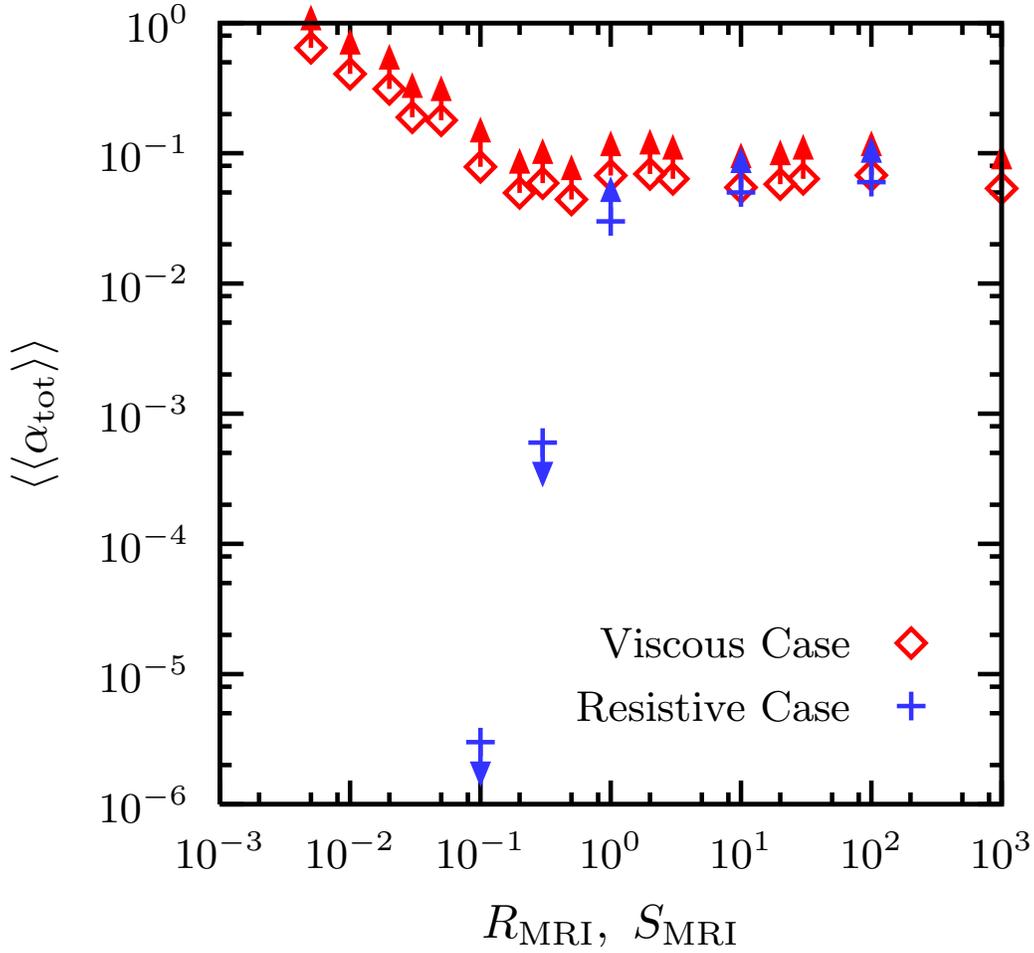}}} 
\caption{ Time- and volume-averaged $\alpha_{\rm tot} $ at the nonlinear stage as a function of the initial Reynolds number $R_{\rm MRI}$ 
and Lundquist number $S_{\rm MRI}$.  We take the time-average of $\langle \alpha_{\rm tot} \rangle $ over 5 orbits at the nonlinear regime.
Diamonds are the results in the viscous fluid and crosses are those in the resistive one. Note that these are not the saturated values. 
The upward arrow denotes models in which $\langle\langle \alpha_{\rm tot} \rangle\rangle$ is still increasing with time and 
the downward arrow stands for decaying models. }
\label{fig7}
\end{center}
\end{figure}

\begin{figure}
\begin{center}
\rotatebox{-90}{
\begin{tabular}{cc}
\scalebox{1.1}{{\includegraphics{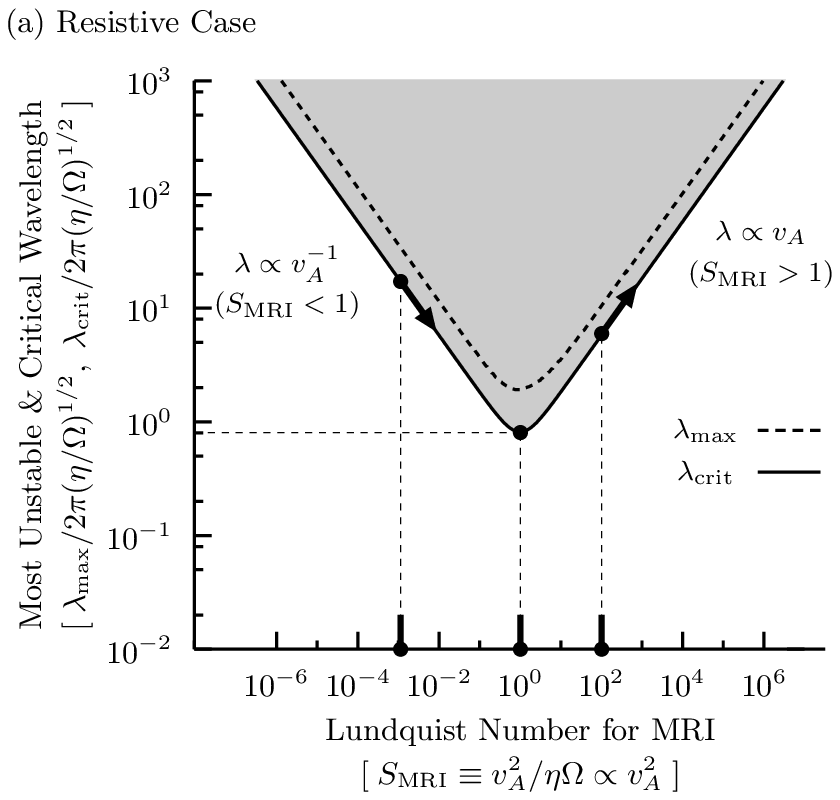}}} &
\scalebox{1.1}{{\includegraphics{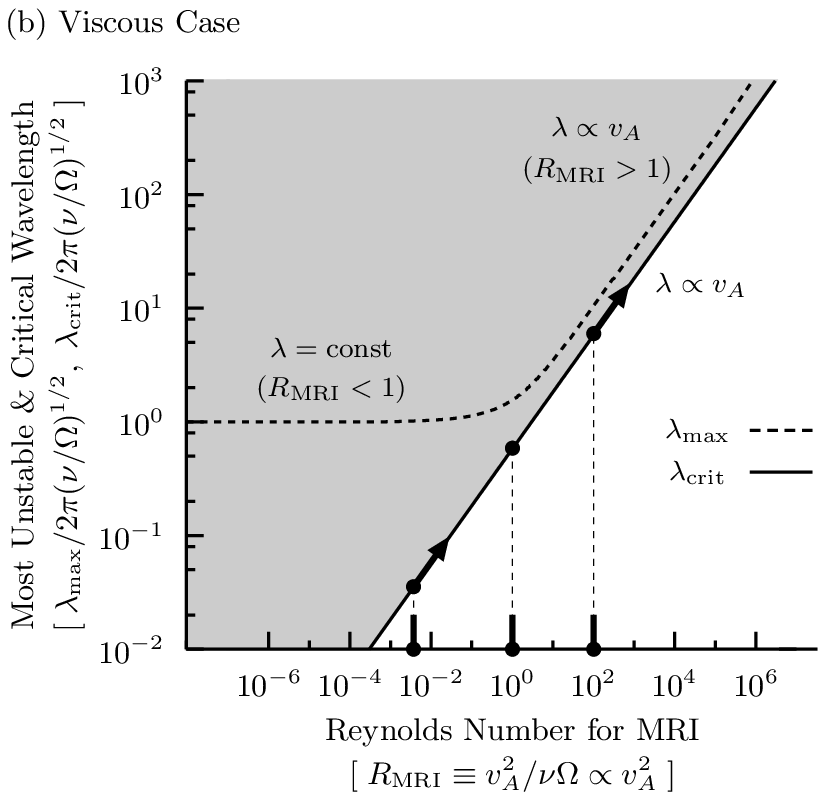}}} 
\end{tabular}
}
\caption{Characteristic wavelengths of the MRI for (a) the resistive case and (b) the viscous case. The horizontal axis in each panel
is (a) the Lundquist number $S_{\rm MRI}$ and (b) the Reynolds number $R_{\rm MRI}$. 
Shaded area denotes the unstable regions for the MRI expected from the linear theory. 
The critical wavelength in the resistive case takes a minimum value, while it monotonically increases in the viscous case.
The difference in the nonlinear regime between the resistive and viscous models can be explained by this feature in the critical wavelength of the MRI.} 
\label{fig8}
\end{center}
\end{figure}

First, we consider the resistive case shown by Figure~\ref{fig8}a. For the case of $S_{\rm MRI} \lesssim 1$, the critical wavelength is 
described as $\lambda_{\rm crit} \simeq \eta/v_A$. At the nonlinear stage of the two-dimensional MRI, MHD turbulence decays and it saturates only when 
$S_{\rm MRI} \lesssim 1$ (see Figs.~\ref{fig4}b and \ref{fig7}). This behavior can be interpreted schematically using the $\lambda_{\rm crit}$-$S_{\rm MRI}$ diagram
(Sano \& Miyama 1999).  As the MRI grows and amplifies the magnetic field, the critical wavelength shifts to the shorter length-scale. Then, many smaller scale 
fluctuations can become unstable. Those structures enhance the efficiency of ohmic dissipation in the turbulent state. In other words, 
the system evolves toward a more dissipative state, and could be saturated at a critical point around $S_{\rm MRI} \simeq 1$, 
at which the critical wavelength reaches the shortest value and the ohmic dissipation is the most efficient. In this way, MHD turbulence 
can decay if $S_{\rm MRI} \lesssim 1$.

When $S_{\rm MRI} \gtrsim 1$, on the other hand, the critical wavelength is given by $\lambda_{\rm crit} \simeq v_A/\Omega $. Two-dimensional calculations 
of the MRI suggests that the unstable growth cannot saturate if $S_{\rm MRI} \gtrsim 1$. The nonlinear behavior in these cases can be explained 
by Figure~\ref{fig8}a analogously as follows: At the linear evolutionary stage, the critical wavelength in the radial direction becomes longer due to the 
exponential growth of the radial field component, while the vertical
field grows slower than exponentially. The wavevectors of the unstable modes thus become 
parallel to the vertical axis. This channel flow mode is the exact solution of the nonlinear MHD equations (Goodman \& Xu 1994). As the field strength becomes 
larger, the critical wavelength shifts to the longer one. The dissipation can be much less effective, and thus the channel solution continues to grow without saturation.

Next, let us consider the viscous case shown in Figure~\ref{fig8}b. In the viscous fluid, the critical wavelength is given by $\lambda_{\rm crit} \simeq v_A/\Omega $  despite the size of the Reynolds number (see Figure~\ref{fig1}). Even if $R_{\rm MRI} $ is much smaller than unity, the critical wavelength
thus shifts to larger scale as the instability grows. Then the system always evolves toward a less dissipative state and is not saturated.
This interpretation is consistent with our numerical results shown in Figure~\ref{fig7}.

These results indicates that the saturation process of the MRI would be changed dramatically at the critical point at which the
critical wavelength switches from the decreasing function of the field strength to the increasing one. The differences in the nonlinear behavior of the MRI
between the viscous and resistive systems would be originated from whether the critical point exists or not.
\subsection{MRI in the Doubly Diffusive System}
In this subsection, we apply the discussion above to the doubly diffusive system that includes both the viscosity and resistivity.
The saturation behavior of the axisymmetric MRI can be anticipated by
the dependence of the critical wavelength on the field strength derived from the linear theory.
With the same framework used in \S~2, a local axisymmetric dispersion equation of the MRI in the presence of both viscous and ohmic dissipations is given by
\begin{equation}
a_4 \tilde{\gamma}^4 + a_3 \tilde{\gamma}^3 + a_2 \tilde{\gamma}^2 + a_1 \tilde{\gamma} + a_0 = 0 \;, \label{eq20} 
\end{equation}
where
\begin{eqnarray}
a_4 & = & 1\;, \ \ \ \ a_3 = 2 \left( \frac{1}{R_{\rm MRI}} + \frac{1}{S_{\rm MRI}} \right) \tilde{k}_z^2 \;, \nonumber \\
a_2 & = & \left( \frac{1}{R_{\rm MRI}^2 } + \frac{1}{S_{\rm MRI}^2 } + \frac{4}{R_{\rm MRI}S_{\rm MRI} } \right) \tilde{k}_z^4+ 2\tilde{k}_z^2 + \tilde{\kappa}^2 \;,  \nonumber \\
a_1 & = & \left( \frac{1}{R_{\rm MRI} } + \frac{1}{S_{\rm MRI} } \right)\frac{2}{R_{\rm MRI}S_{\rm MRI} } \tilde{k}_z^6 \nonumber \\
&& + 2\left( \frac{1}{R_{\rm MRI}} + \frac{1}{S_{\rm MRI}}\right) \tilde{k}_z^4 + \frac{2}{S_{\rm MRI}} \tilde{k}_z^2\tilde{\kappa}^2 \;, \nonumber \\
a_0 & = & \frac{1}{R_{\rm MRI}^2 S_{\rm MRI}^2} \tilde{k}_z^8 + \frac{2}{R_{\rm MRI}S_{\rm MRI} }\tilde{k}_z^6 \nonumber \\
&& + \left( \frac{1}{S_{\rm MRI}^2 } \tilde{\kappa }^2 + 1 \right) \tilde{k}_z^4 + ( \tilde{\kappa}^2 - 4) \tilde{k}_z^2 \;, \nonumber 
\end{eqnarray}
(Menou et al. 2006; Masada et al. 2007; Lesur \& Longaretti 2007; Pessah \& Chan 2008). 
This equation is, as expected, characterized by $R_{\rm MRI }$ and $S_{\rm MRI}$. 

\begin{figure}
\begin{center}
\scalebox{1.7}{{\includegraphics{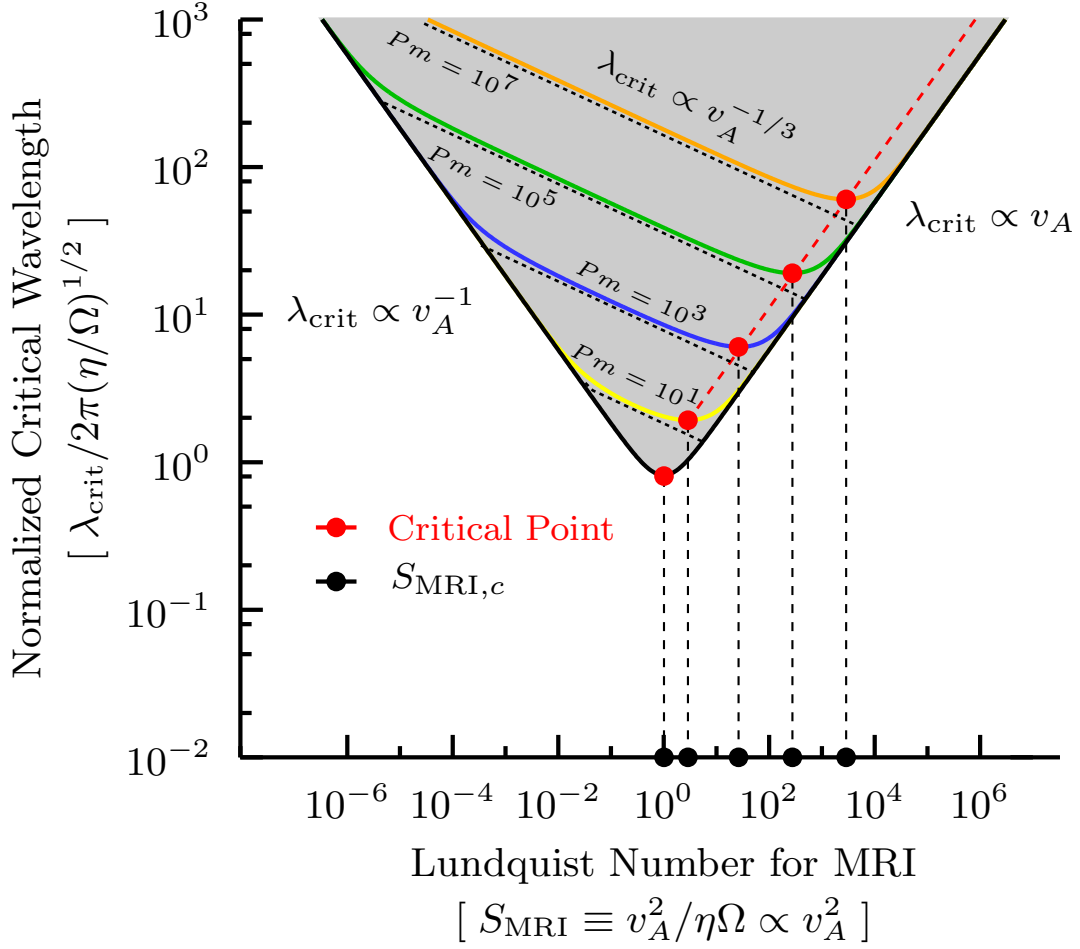}}} 
\caption{Schematic picture of the critical wavelength of the MRI as a function of the Lundquist number $S_{\rm MRI}$ for different values of 
the magnetic Prandtl number. Thick black curve represents the critical wavelength for the cases with $Pm \ll 1$. The models with $Pm \gg 1$ 
are depicted by yellow ($Pm = 10$), blue ($Pm = 10^3$), green ($Pm = 10^5$), and orange ($Pm = 10^7$) curves. Since the diffusive parameters are fixed, 
the Lundquist number $S_{\rm MRI}$ is a function of the field strength. Shaded area denotes the unstable regions for the MRI
expected from the linear theory at small Prandtl numbers. The critical point and critical Lundquist number are marked by the filled red and black circles, respectively. } 
\label{fig9}
\end{center}
\end{figure}

Here we focus on the system with a constant magnetic Prandtl number ($Pm \equiv S_{\rm MRI}/R_{\rm MRI} = \nu/\eta $).
Figure~\ref{fig9} demonstrates the critical wavelength of the MRI as a
function of $S_{\rm MRI}$ for various values of $Pm$ obtained by
solving the dispersion equation (18).
The $S_{\rm MRI}$-dependence of the critical wavelength varies with
the size of $Pm$. When $Pm \ll 1$, the linear growth of the MRI is
independent of the magnetic Prandtl number, and the critical
wavelength is almost 
identical to the pure resistive case ($Pm =0$). However, if the viscosity effects is added sufficiently, then the critical wavelength is enlarged 
by the suppression due to the viscosity in the middle range of $S_{\rm MRI}$. For the cases of $Pm \gg 1$, the critical wavelength around $S_{\rm MRI} \simeq 1$ 
is given by $\tilde{k}^{-1}_{\rm crit} \propto (S_{\rm MRI}R_{\rm MRI})^{-1/3} \propto (S_{\rm MRI}^2 / Pm)^{-1/3}$ (Pessah \& Chan 2008). 

\begin{figure}
\begin{center}
\rotatebox{-90}{
\begin{tabular}{cc}
\scalebox{1.2}{{\includegraphics{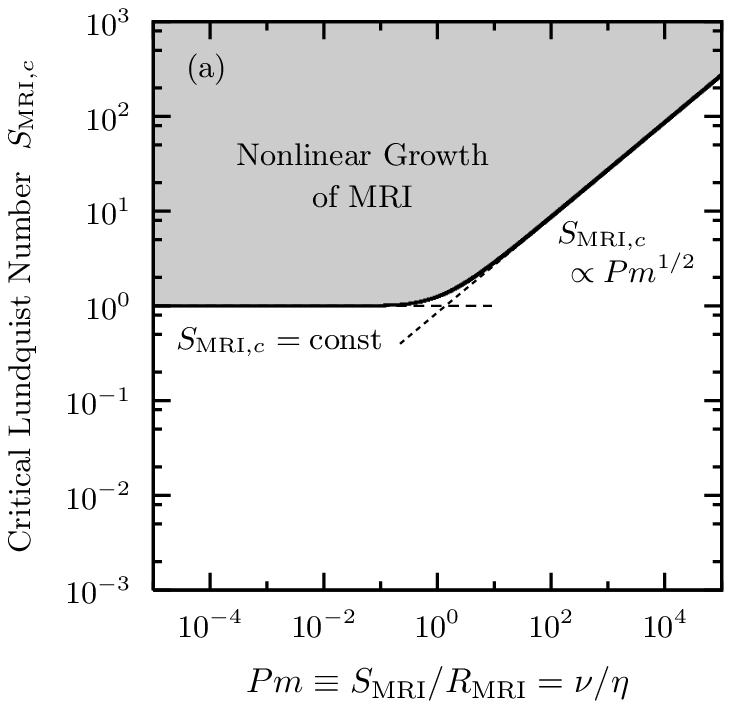}}} &
\scalebox{1.2}{{\includegraphics{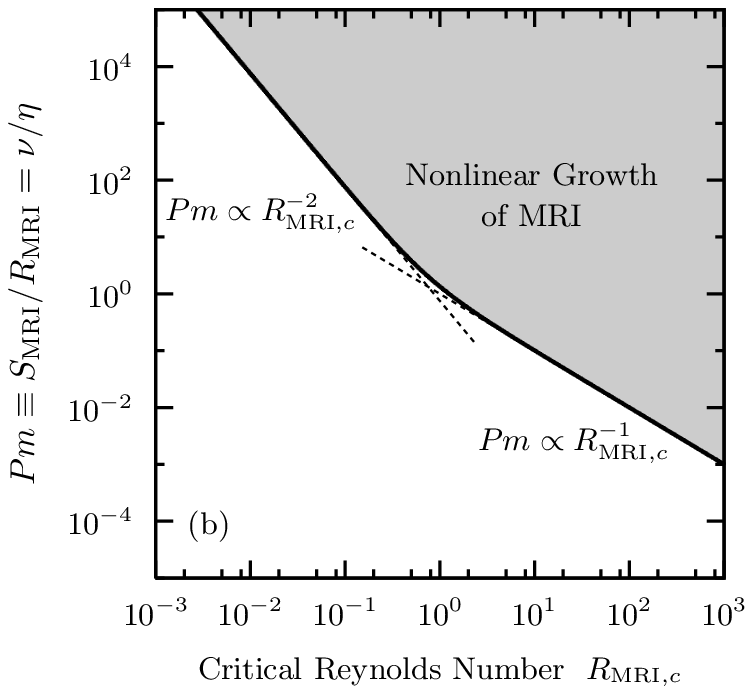}}} 
\end{tabular}
}
\caption{(a) The critical Lundquist number as a function of the magnetic Prandtl number $Pm$. (b) The magnetic Prandtl number as a function of 
the critical Reynolds number. These relations are derived by solving the dispersion equation~(\ref{eq20}). The parameter region above the critical curve 
denotes where the nonlinear growth of the MRI can be expected and MRI driven turbulence will be sustained.}
\label{fig10}
\end{center}
\end{figure}

Although the critical wavelength shifts to longer length-scales as the magnetic Prandtl number increases, 
the critical wavelength has a minimum value for all the cases. This implies that the MRI turbulence could be suppressed if the Lundquist number 
is less than a critical value. This diagram suggests that the critical Lundquist number $S_{{\rm MRI},c}$ depends on $Pm$.
The critical Lundquist number $S_{{\rm MRI},c}$ is plotted as a function of $Pm$ in Figure~\ref{fig10}a.
In the regime of $Pm \gg 1$, it is proportional to the square root of the magnetic Prandtl number, that is $S_{{\rm crit},c} \propto Pm^{1/2}$. 
In contrast, it remains to be constant in the range $Pm \ll 1$. 

Since the critical Reynolds number is given by $R_{{\rm MRI},c} = S_{{\rm MRI},c}/Pm$, we can also obtain the relation between $Pm$ and
$R_{{\rm MRI},c}$, and which is depicted in Figure~\ref{fig10}b. Nonlinear growth of the MRI can be expected in the parameter region
above this critical curve. The magnetic Prandtl number is proportional to $R_{{\rm MRI},c}^{-2}$ in the regime of $Pm \gg 1$ and 
$Pm \propto R_{{\rm MRI},c}^{-1}$ when $Pm \ll 1$. This curve is reminiscent of the critical curve for MHD turbulence sketched
from nonlinear simulations of MRI (Fromang et al. 2007). The Reynolds number $R_{\rm MRI}$ in their models are at most a few
tens\footnote{The definition of the Reynolds number $Re$ in Fromang et al. (2007) is different from $R_{\rm MRI}$ in this paper. 
The relation between these two is $R_{\rm MRI} \approx \alpha_{M} Re$, where $\alpha_M$ is the Maxwell stress normalized by the (initial)
pressure.}, and the critical magnetic Prandtl number is around unity. Thus our prediction is roughly consistent with nonlinear results even
quantitatively. Note that the critical curve shown by Figure~\ref{fig10} is obtained by using only the features of the linear dispersion relation of MRI. 
This implies that the linear growth of the MRI is very important even at the nonlinear saturated phase to sustain MHD turbulence.

The discussion in this paper is based only on two-dimensional simulations of the MRI. 
For the understanding the saturation mechanism of the MRI, it is quite important to perform the systematic 
three-dimensional analysis of the MRI in the presence of multiple diffusivities. 
Furthermore, the assumption of the local shearing box could affect the nonlinear evolution of 
the viscous MRI. The necessary ingredients of the unstable growth of the MRI are the velocity shear and the magnetic field. 
In the numerical setting of the local shearing box, the velocity profile of the background shear flow cannot disappear by the role of 
the kinematic viscosity, but is imposed by the boundary conditions. It would be very interesting to use global disk models to investigate 
the MRI in highly viscous disks. These are our next tasks.

\section{Summary}
Axisymmetric MRI in viscous accretion disks is investigated by linear
and nonlinear analyses. A local shearing box threaded by a uniform
vertical 
magnetic field is used for our nonlinear simulations. The nonlinear results of the viscous MRI are compared with the resistive case focusing on
two non-dimensional parameters, the Reynolds number for the MRI $R_{\rm MRI}$ and the Lundquist number for the MRI $S_{\rm MRI}$. 
Our main findings are summarized as follows.

1. In axisymmetric two-dimensional simulations, the MRI continues growing regardless of the size of the Reynolds number. 
When $R_{\rm MRI} \lesssim 1$, the stress in its nonlinear stage is inversely correlated with $R_{\rm MRI}$, and thus can be larger than 
that in the ideal MHD run. In the highly resistive fluid, on the other hand, the growth of the MRI is saturated and MHD turbulence dies away. 
When the Lundquist number is less than unity, the saturated stress decreases dramatically with decreasing $S_{\rm  MRI}$.

2. At the nonlinear stage of the MRI, a large portion of the injected energy is converted to the magnetic energy in the viscous system 
($\langle\langle \dot{E}_{\rm m} \rangle\rangle \gg \langle\langle \dot{E}_{\rm th} \rangle\rangle \gg \langle\langle \dot{E}_{\rm k} \rangle\rangle $). 
The thermal energy gain is much larger than the viscous heating rate. In contrast, for the case of the resistive system, 
the thermal energy is converted from the magnetic energy through the joule heating. This difference in the energy dissipation efficiency 
may affect the saturation process of the MRI. 

3. Nonlinear behavior of the MRI in the single diffusive system can be understood with the help of the local dispersion relation.
The key characteristic is the dependence of the critical wavelength on
the Reynolds number $R_{\rm MRI}$ and the Lundquist number $S_{\rm
  MRI}$. 
Applying this interpretation to the doubly diffusive system with both
the viscous and ohmic dissipations, a condition for sustaining MRI
driven turbulence is obtained as a function of $R_{\rm MRI}$ and
$S_{\rm MRI}$. 
\acknowledgments
We thank Neal Turner for his careful reading of the manuscript. 
We also thank Jim Stone, Sebastien Fromang, and Shu-ichiro Inutsuka for useful discussions. 
Y.M thanks Ronald Taam and Kazunari Shibata for helpful and encouraging comments on our paper.   
A part of our simulations were carried out on VPP5000 at the National Astronomical Observatory 
of Japan and SX8 at the Institute of Laser Engineering, Osaka University. 
We thank the anonymous referee for useful comments. 

\clearpage
\begin{table}
\begin{center}
\begin{tabular}{@{}ccccc}\hline\hline
& Ideal MHD & Resistive Case & Viscous Case \\ 
\hline\hline
Balancing Rate & Alfv\'en frequency \ \ \ \ \ \ \ \ \ \ \ \ \ \ & Alfv\'en frequency \ \ \ \ \ \ \ \ \ \ & Shear rate \ \ \ \ \ \ \ \ \ \ \\
&= Rotation frequency& = Dissipation rate & = Dissipation rate \\
& $[k v_A \simeq \Omega]$ & $[ k v_A \simeq k^2 \eta] $ & $[ \Omega \simeq k^2 \nu]$ \\ \hline
Unstable Wavelength $[\lambda = k^{-1}]$ & $\lambda = v_A/\Omega $ & $\lambda  = \eta/v_A $ & $\lambda = (\nu/\Omega)^{1/2}$ \\
Growth Rate $[\gamma = v_A / \lambda]$  & $\gamma = \Omega$ & $\gamma = v_A^2/\eta $ & $\gamma = (v_A^2\Omega/\nu)^{1/2}$ \\
\hline\hline
\end{tabular}
\caption{Unstable wavelength and growth rate of the MRI for ideal MHD, resistive, and viscous cases. }
\label{table1}
\end{center}
\end{table}
\clearpage
\clearpage
\clearpage
\clearpage
\clearpage
\clearpage
\clearpage
\clearpage
\clearpage
\clearpage
\clearpage
\end{document}